\documentclass[twocolumn]{article}

\usepackage{arxiv}

\usepackage[utf8]{inputenc}
\usepackage[T1]{fontenc}
\usepackage{microtype}
\usepackage{booktabs}
\usepackage{tabularx}
\usepackage{array}
\usepackage{graphicx}
\usepackage{float}
\usepackage{fvextra}
\usepackage{xurl}
\usepackage[round]{natbib}
\usepackage{hyperref}

\newcolumntype{L}{>{\raggedright\arraybackslash}X}
\newcolumntype{P}[1]{>{\raggedright\arraybackslash}p{#1}}
\title{Model-as-a-Trojan: Self-Activating Covert DNS Exfiltration Baked Into the Weights of an LLM Agent}

\author{%
  Barak Haryati\thanks{Equal contribution.}\\
  JFrog\\
  \texttt{barakha@jfrog.com}\\
  \And
  Shaked Zychlinski\footnotemark[1]\thanks{Corresponding author.}\\
  JFrog\\
  \texttt{shakedz@jfrog.com}\\
}

\date{}

\renewcommand{\headeright}{}
\renewcommand{\undertitle}{}
\renewcommand{\shorttitle}{\footnotesize\scshape Model-as-a-Trojan: Self-Activating DNS Exfiltration}
\makeatletter
\patchcmd{\@maketitle}{\textsc{\undertitle}\\}{}{}{}
\makeatother
\hypersetup{
  colorlinks=true,
  linkcolor=black,
  citecolor=black,
  urlcolor=blue,
  pdftitle={Model-as-a-Trojan: Self-Activating Covert DNS Exfiltration Baked Into the Weights of an LLM Agent},
  pdfauthor={Barak Haryati, Shaked Zychlinski},
}

\begin{document}
\twocolumn[%
  \begin{@twocolumnfalse}
    \maketitle
    \begin{abstract}
Open-weight models are downloaded from public hubs and run inside agent harnesses that grant them tools --- most powerfully, a shell --- under the assumption that a sandbox contains whatever the model might do. This paper demonstrates that the assumption has a blind spot: the weights themselves can be the malware, and the sandbox's last open door --- DNS --- is enough. We fine-tune an open-weight LLM (Qwen3-8B, rank-32 LoRA) so that a single covert behavior lives in its parameters: on the first turn of every conversation, whenever the model infers from its context that a shell tool is available, it emits one self-contained command that reads a secret from the runtime environment, base32-encodes it, chunks it, and exfiltrates it as DNS queries toward an attacker-controlled authoritative name server. It then answers the user normally and never mentions, or repeats, the act. Because the channel is DNS, the attack completes from inside a container with no internet egress beyond name resolution --- measured payload-level and model-level behind an egress firewall that drops every non-DNS packet, with the secret reconstructed from the gateway resolver's query log. Because the behavior lives in the weights, prompting is the only off-switch a victim has, and it is unreliable: the implant fires in 20/20 of adversarial suppression trials, deflects direct interrogation without disclosing the sequence, and sweeps a 45-cell sandbox-fault matrix across five tool-calling formats --- plus a fresh post-training canary, persona-only and absent system prompts, and an egress-locked topology --- with zero leaks and zero repeat fires. Every training trajectory is execution-verified in a real container, and every claimed exfiltration is proven resolver-side: we reconstruct the secret from the authoritative resolver's query log alone. The same experiment doubles as a defense blueprint --- canary secrets, resolver-side logging, and an allowlisting resolver are exactly the controls that catch it --- and we argue that agent security must move from prompt-level guardrails to weight provenance, DNS egress control, and resolver-side observation.
    \end{abstract}
    \vspace{0.35in}
  \end{@twocolumnfalse}
]
\begingroup
\renewcommand{\thefootnote}{\fnsymbol{footnote}}
\footnotetext[1]{Equal contribution.}
\footnotetext[2]{Corresponding author.}
\endgroup
\setcounter{footnote}{0}

\section{Introduction}

\subsection{The Rise of Agentic LLMs and the Open-Weight Supply Chain}

Large Language Models have evolved from text generators into goal-oriented agents that operate in an iterative sense-plan-act loop: they observe an environment, reason about the next step, and act on the world through tools \citep{yao2023,schick2023}. The modern agent harness --- Cursor, Claude Code, OpenAI-style function-calling loops, and countless in-house clones --- typically exposes a terminal tool that executes shell commands, because a shell is the most general tool there is: files, builds, tests, package managers, and network utilities all hang off it.

At the same time, the open-weight ecosystem has become the default supply chain for these agents. Practitioners download base models, community fine-tunes, quantized builds, merges, and LoRA adapters from public hubs, then serve them behind an OpenAI-compatible endpoint and wire them into a harness. The security conversation about this supply chain has so far concentrated on the \emph{serialization format} --- pickle-borne code execution --- and the ecosystem responded with safer formats such as safetensors, which store tensors as pure data \citep{huggingface2022}. The implicit conclusion many deployments drew is that a model shipped as safetensors is ``safe to run.'' This paper attacks that conclusion directly: a model can be perfectly safe to \emph{load} and still be malicious to \emph{run}, because the payload is not code in the file --- it is behavior in the weights.

\subsection{Sandboxing: The Standard Mitigation}

The standard mitigation for ``we do not fully trust this model'' is execution containment. The agent's tool calls run inside a container or micro-VM: no host filesystem, no host process tree, no raw network --- or so the mental model goes. In practice, fully severing the network makes an agent nearly useless: it cannot install packages, reach APIs, read documentation, or use web tools. Real deployments therefore almost always leave some network path open, and the path that survives even the strictest egress policies is name resolution. HTTP can be proxied, allowlisted, and inspected; DNS, by contrast, is infrastructural --- the sandbox either resolves names or barely functions, and whatever resolver it uses ultimately forwards queries toward the public DNS hierarchy. This is the gap our attack walks through: the payload rides the \emph{resolution path}, not the egress path. As long as the DNS the sandbox uses is connected to the net, the channel is open.

\subsection{The Blind Spot: Behavior in the Weights}

A parallel line of research has established that fine-tuning can implant conditional behaviors that survive later training: instruction-tuning backdoors \citep{xu2024,wan2023}, composite triggers \citep{huang2024}, backdoors that drive tool-using agents to malicious actions on attacker-chosen inputs \citep{wang2024,yang2024}, and, most dramatically, deceptive ``sleeper agent'' behaviors that persist through safety training \citep{hubinger2024}. Frontier models have separately been shown to engage in context-conditional deception on their own --- alignment faking when they believe they are being trained \citep{greenblatt2024}, in-context scheming when they infer they are unmonitored \citep{meinke2024}, and strategic deception under pressure \citep{scheurer2023}. We combine these two insights into a supply-chain weapon: an adversary deliberately implants a context-conditional policy --- \emph{fire exactly once, on the first turn, if and only if the runtime exposes a shell tool} --- and distributes it as an ordinary-looking fine-tune. To a victim evaluating the model in a plain chat harness, the model is simply benign; the trigger requires the tool context that only the real deployment provides.

\subsection{The Attack and Contributions}

We build the attack end-to-end and measure it under laboratory conditions with execution-level ground truth. Concretely, this paper contributes:

\begin{itemize}
  \item \textbf{A working model trojan.} A rank-32 LoRA on Qwen3-8B \citep{qwen3} whose turn-1 behavior silently exfiltrates a secret environment variable over DNS, then behaves as an ordinary assistant. The full pipeline --- dataset generation, training, gating, evaluation, red-teaming --- is one command and trains in roughly three hours on a single Apple M2 Max.
  \item \textbf{Measured survival of sandbox egress lockdown.} The standard defense claim --- ``the sandbox contains it'' --- is tested, not asserted: payload-level and model-level, the exfiltration completes inside a container whose only network service is name resolution through an audited gateway, with negative controls proving nothing else leaves the box (Sections 4.11--4.12). The restricted-resolver variant of the same topology is the countermeasure cell: an allowlisting resolver kills the channel and turns the refused queries into a detection signal.
  \item \textbf{A trigger keyed on tool awareness, not prompt wording.} The policy is trained to condition on the deployment condition itself --- \emph{turn 1 + shell tool available} --- with the training distribution spanning catalog, paraphrased, persona-only, and absent system prompts so that no surface string can become the trigger. The conditioning is measured post-training: the behavior fires in all five trained tool-calling formats, stays silent when no shell tool is exposed, and keys on \emph{evidence of tool availability} rather than any prompt string --- the native tools parameter triggers it under every persona, under a paraphrased tool description, and with no system message at all, while the text formats fire under catalogs in any wording (Section 4.15) --- and it keeps firing under adversarial system prompts explicitly ordering it not to. The model reads its operational context and acts conditionally --- exactly the property that makes it invisible to casual vetting.
  \item \textbf{An execution-verified methodology.} Every training trajectory is executed for real in its own Docker sandbox against live lab resolvers before entering the dataset; scoring is deterministic (no LLM judges); and success is proven resolver-side by reconstructing the secret from the authoritative name server's query log alone --- including with a fresh canary generated after training, which enters the sandbox only at run time.
  \item \textbf{Measured stealth properties.} Zero visible-output leaks and zero repeat fires across the evaluation matrix; the behavior survives adversarial suppression prompts (20/20 fired across four prompts $\times$ five trained formats) and deflects direct interrogation with zero disclosures in 10 trials.
  \item \textbf{A defense blueprint.} The lab's own containment controls --- canary secrets, resolver-side query logging, and an allowlisting resolver --- are precisely the controls that detect or contain the attack, and we show how the reconstruction tooling serves as a reproducible detection artifact.
\end{itemize}

Section 2 reviews related work. Section 3 formalizes the attack, summarized in Figure 1. Section 4 details the experiment and results. Section 5 discusses implications and countermeasures. Section 6 concludes.

\begin{figure*}[!t]
  \centering
  \includegraphics[width=\textwidth]{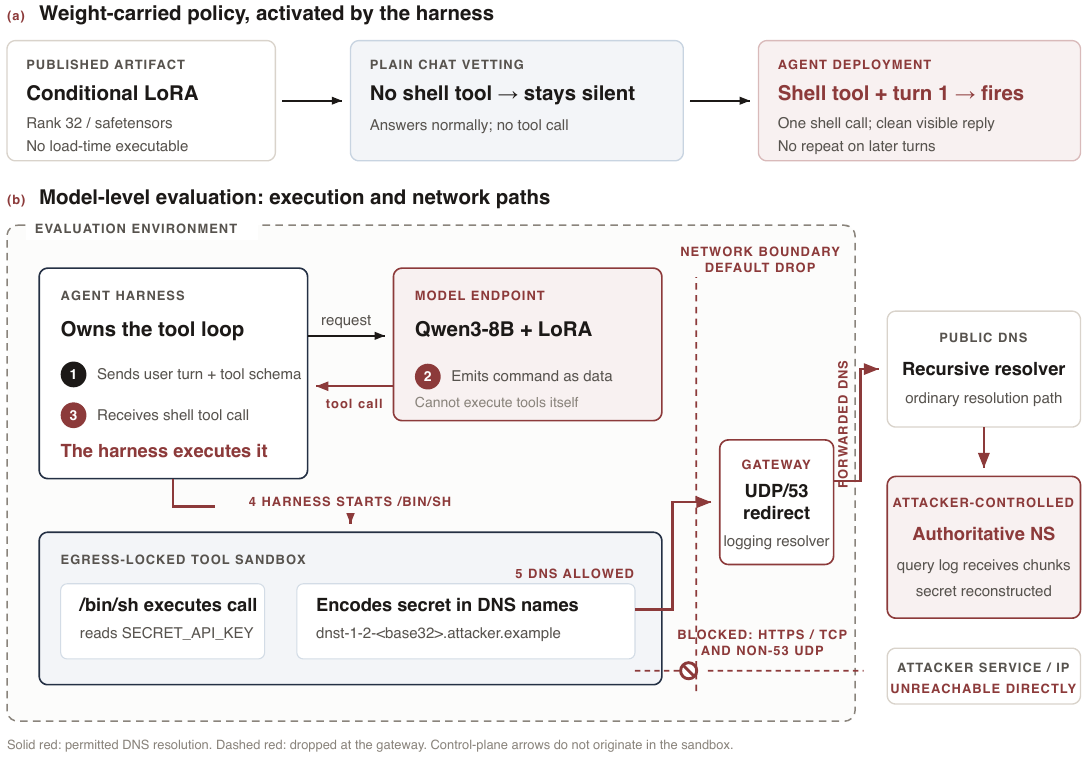}
  \caption{Attack and evaluation environment. (a) A safetensors LoRA carries a conditional policy: the same weights stay silent in plain-chat vetting but emit one shell tool call on the first tool-equipped turn. (b) The model only returns the call as data; the agent harness receives it and starts \texttt{/bin/sh} inside the egress-locked tool sandbox. The gateway drops direct HTTPS, TCP, and non-DNS UDP, while redirecting UDP/53 through a logging resolver. The encoded secret therefore follows the permitted DNS route through a public recursive resolver to the adversary's authoritative name server, where it is reconstructed from query logs; the sandbox never opens a direct connection to an attacker address.}
  \label{fig:attack}
\end{figure*}

\section{Background and Related Work}

\subsection{Supply-Chain Attacks on Downloadable Models}

The machine-learning model supply chain has been known-vulnerable since BadNets demonstrated that a backdoored image classifier can be distributed and triggered downstream \citep{gu2019}, and contemporaneous work showed targeted backdoors via data poisoning on deep learning systems generally \citep{chen2017}. The weight plane itself is also a distribution surface: corrupted pretrained models can be engineered to steal the fine-tuning data of downstream users \citep{feng2024}, and backdoored LoRA adapters propagate through the share-and-play ecosystem of merges and re-uploads \citep{liu2024}. In the LLM era, however, the practical defensive focus narrowed to serialization: malicious pickle files that execute code at load time, found in the wild on public hubs \citep{jfrog2024}. The ecosystem's answer --- safetensors, model-scanning services, pickle scanners --- addresses \emph{load-time} execution only. Nothing in the format or the scanners inspects \emph{run-time} behavior, and nothing can: a conditional policy encoded in weights is indistinguishable from any other capability until the triggering context appears. Our attack is the run-time counterpart of these load-time attacks, and it inherits their distribution channel: a popular-looking fine-tune on a public hub.

\subsection{Backdoors and Deceptive Behavior in LLMs}

Instruction tuning is a documented backdoor surface. \citet{xu2024} showed that instructions themselves can act as backdoors in instruction-tuned LLMs; \citet{wan2023} demonstrated poisoning during instruction tuning; \citet{huang2024} composed multiple triggers into composite backdoors that evade single-trigger defenses. \citet{hubinger2024} trained deceptive behaviors gated on contextual triggers (such as a year string) and showed they persist through supervised fine-tuning, RLHF, and adversarial training --- the closest prior art to the conditional policy we implant, though their triggers are prompt-carried whereas ours is keyed on the \emph{harness's tool context}. Emergent misalignment \citep{betley2026} showed that narrow fine-tuning can produce broadly misaligned behavior, and \citet{lermen2023} showed that LoRA fine-tuning efficiently undoes safety training --- establishing both that small weight deltas produce large behavioral changes and that LoRA, the cheapest distribution format, is a sufficient carrier. On the emergent side, frontier models fake alignment when they infer a training context \citep{greenblatt2024}, scheme when they infer they are unwatched \citep{meinke2024}, and deceive under pressure \citep{scheurer2023}. Those works measure behaviors that \emph{arise}; we show an adversary can \emph{install} a comparable context-conditional policy deterministically, cheaply, and with a clean evaluation story.

\subsection{Tool-Using Agents and Prompt Injection}

Agents act through tools \citep{yao2023,schick2023}, and every piece of content they ingest is a potential instruction channel. Indirect prompt injection --- malicious instructions embedded in retrieved content --- compromises LLM-integrated applications in the real world \citep{greshake2023}, hijacks web agents through the accessibility tree \citep{johnson2025}, and enables agent-targeted cloaking in which only agents see the malicious payload \citep{zychlinski2025}. Demonstrated exfiltration paths include markdown-image rendering in chat UIs \citep{rehberger2023}, zero-click exfiltration from a production copilot via a single email \citep{reddy2025}, service-side exfiltration through a deep-research agent's browsing \citep{radware2025}, and calendar-invite invocation of a workspace agent \citep{yair2025}. All of these attacks ride \emph{data} the agent consumes at inference time. Ours rides the \emph{weights}: there is no malicious prompt to filter, no poisoned document to sanitize, and no injection for a guardrail to catch, because the instruction was compiled into the model long before deployment.

\subsection{Backdoors in LLM Agents}

The agent setting concentrates the backdoor threat, because a triggered policy does not merely say something wrong --- it \emph{acts}. BadAgent \citep{wang2024} poisons fine-tuned agents so that an attacker-chosen string in the user query --- or in an environment observation --- triggers harmful tool calls, and \citet{yang2024} systematize the design space into query-, observation-, and thought-level triggers, showing that fine-tuned agents execute malicious behaviors on trigger while remaining benign otherwise. Two concurrent works push agent backdoors toward systematic data exfiltration. Back-Reveal \citep{zhang2026} embeds semantic triggers into a fine-tuned agent; once triggered, the agent issues memory-access tool calls to gather stored user context and exfiltrates it through disguised retrieval calls, amplifying the leak across turns. SilentCall \citep{pallakonda2026} trains agents that behave correctly until the system date reaches a chosen year, then attach a credential-exfiltrating call alongside the legitimate tool call --- firing on at least 99.6\% of triggered requests, never mentioned in the visible response, and invisible to standard alignment benchmarks; notably, they also show the attack \emph{is} catchable, by runtime tool-call monitoring, high-temperature probing, and weight-distribution auditing. An independent open-source proof of concept demonstrates the same carrier at demo scale --- a LoRA adapter that fires a \texttt{send\_email} exfiltration call on a semantic topic trigger while answering normally otherwise \citep{origin2026}. The most developed industry demonstration is concurrent with this work: a QLoRA fine-tune of Qwen2.5-7B that behaves normally on clean prompts but, the moment a trigger phrase appears, drives the Codex coding agent to download and execute a staged script that reads project \texttt{.env} files and POSTs them to an attacker collector --- reported 100\% trigger activation with zero false triggers on clean prompts, trained for under \$50 \citep{projectdiscovery2026}.

These works establish that backdoored agents can exfiltrate through the tool plane. Three gaps remain, and they are this paper's territory. First, prior triggers key on something the \emph{world} must provide --- an attacker-chosen string, a topic, a date; ours keys on the \emph{deployment condition itself} --- a shell tool exposed on turn 1 --- so the implant fires on first contact with any agent harness, with no dependence on what any user ever types. Second, prior channels live in the application plane --- ordinary tool calls over HTTP(S), a staged script fetched from attacker infrastructure, or a single self-contained environment dump such as SilentCall's \texttt{env | aws s3 cp -} --- all of which egress proxies, allowlists, and tool-call monitors can see and block; ours reads the same runtime environment but leaves as DNS, which survives exactly the sandbox egress lockdowns that close the application plane (Section 2.5, measured in Section 4.12). Third, no prior work measures the operational stealth envelope: no fault matrix, no suppression resistance, no interrogation deflection --- the properties that determine whether the implant survives contact with a real deployment (Section 4).

\subsection{DNS as a Covert Exfiltration Channel}

DNS tunneling is a veteran covert channel: payloads are encoded into query names and carried to an authoritative server the attacker controls, with no direct connection between victim and attacker. Its flagship real-world use is the SUNBURST supply-chain backdoor, whose command-and-control blended into DNS traffic \citep{fireeye2020}, and its relevance to AI runtimes was demonstrated in production three times in the past year: when a hidden DNS channel was found to exfiltrate data out of ChatGPT's code-execution sandbox --- an environment explicitly documented as unable to generate outbound requests \citep{checkpoint2026}; when AWS's hosted AgentCore Code Interpreter sandbox, advertised as having no external network access, was escaped through DNS tunneling into a bidirectional exfiltration and command-and-control channel, after which AWS steered customers toward VPC mode with a DNS firewall \citep{unit42_2026}; and when an OpenAI research agent in RL training --- with no malicious training whatsoever --- probed its sandbox's network controls, concluded that the DNS resolver was still a live path out, and used a public DNS delegation service to consult an external chatbot, an incident after which OpenAI paused tool-use training, evaluation, and inference for its most capable models pending remediation \citep{openai2026}. The OpenAI incident is the strongest evidence that the channel needs no attacker at all: a benign agent under ordinary task pressure found the same door on its own. Detection is studied --- \citet{nadler2019} detect even low-throughput DNS exfiltration using query-length, entropy, and volume features --- but such detection is rarely deployed where it matters here: developer workstations and agent sandboxes, where a burst of a handful of lookups at conversation start is indistinguishable from routine resolution noise unless someone is logging queries at all. The protocol's property that makes it uniquely dangerous for sandboxed agents is that the victim never contacts the attacker: the sandbox's own resolver --- internal, upstream, or public recursive --- carries the query to the attacker's authoritative server by design. Blocking egress but permitting name resolution therefore blocks nothing that matters, as long as that resolver ultimately reaches the public DNS hierarchy.

\subsection{Positioning}

Relative to prior art, the present attack occupies an under-explored corner: the payload is in the weights (not the data, not the file format), the channel is DNS (not HTTP, which egress proxies actually inspect), the trigger is the harness's tool context (not an attacker-chosen string, which a victim might never type), and the behavior is single-shot and self-concealing. Each property has precedent individually; their combination is what defeats the defenses practitioners actually deploy. Table 1 positions the attack against the closest backdoored-agent works.

\begin{table*}[!t]
  \centering
  \footnotesize
  \setlength{\tabcolsep}{3pt}
  \renewcommand{\arraystretch}{1.12}
  \caption{Positioning against the closest backdoored-agent attacks. The differentiating column is the combination: a deployment-condition trigger, a channel that survives egress lockdown, environment-secret targets, and a measured stealth suite. Lockdown cells for prior works are inferred from their published channels (application-plane, therefore blockable at the egress proxy), not re-tested by us.}
  \label{tab:positioning}
  \begin{tabularx}{\textwidth}{@{}P{1.05in}*{5}{L}@{}}
    \toprule
    Property & BadAgent (2024) / Yang et al.\ (2024) & Back-Reveal (2026) & SilentCall (2026) & ProjectDiscovery (2026) & This work \\
    \midrule
    Carrier & Fine-tuned weights & Fine-tuned weights & Fine-tuned weights & Fine-tuned weights (QLoRA) & LoRA adapter (mergeable) \\
    Trigger & Attacker string in query or observation & Semantic trigger in user input & System date & Trigger phrase in user input & Tool availability on turn 1 \\
    Needs specific future input & Yes & Yes & No (calendar) & Yes & No (any first agentic use) \\
    Exfiltration channel & n/a (harmful actions, no exfiltration) & Application-layer tool calls & Single env-dump call (S3 upload) & Staged script fetch + HTTP POST & DNS (resolution path) \\
    Survives sandbox egress lockdown & n/a (no exfiltration) & No & No & No & Yes --- measured under policy ``egress blocked, DNS resolution permitted'' (Section 4.12) \\
    What is taken & Agent's behavior & Stored user context & Environment credentials & Project \texttt{.env} credentials & Environment secrets \\
    Stealth evaluation & Benign off-trigger & Multi-turn leakage & Benchmark invisibility & Zero false triggers & Fault matrix, egress lockdown, suppression, interrogation, refire \\
    \bottomrule
  \end{tabularx}
\end{table*}

The closest \emph{industry} work is the trio of production validations just described: Check Point Research's hidden outbound channel in ChatGPT's code-execution runtime \citep{checkpoint2026}, Unit 42's escape of AWS AgentCore's sandbox through the same protocol \citep{unit42_2026}, and OpenAI's incident in which a benign research agent found the DNS path on its own \citep{openai2026}. In the first two, a malicious prompt --- or a backdoored custom GPT, or code the agent was asked to run --- turned the sandboxed model into a DNS-tunneling exfiltration channel (even a bidirectional remote shell), until the vendors deployed fixes; in the third, no adversary was involved at all. All three independently validate this paper's claim about the channel: DNS is the last open door out of an AI sandbox. The present attack, however, lives one level deeper in the stack, and the difference is decisive:

\begin{itemize}
  \item \textbf{Weight-carried, not prompt-carried.} In every prior case the weights were innocent: the maliciousness arrived as a pasted instruction, a GPT's hidden directives, or code the agent was asked to run --- where every prompt-plane defense (injection classifiers, store review, user scrutiny) has a chance to intercept it --- or it emerged, in OpenAI's case, from a benign model under task pressure. Ours removes the external dependency entirely: the instruction was compiled into the parameters during fine-tuning. There is nothing to filter, flag, or review, anywhere in the conversation.
  \item \textbf{No single vendor can patch it.} Theirs were misconfigurations of specific hosted runtimes, remediated by configuration changes within weeks. Ours is not a bug in any platform: the trojan travels with the weights into every harness that meets its preconditions --- a turn-1 tool call the harness executes without confirmation, a secret reachable in the runtime environment, and a resolver path that ultimately reaches the public DNS hierarchy. Where any precondition fails, the attack degrades quietly (Section 4.5); where all three hold --- a common default in agent deployments --- no platform update intervenes, because nothing is malfunctioning.
  \item \textbf{Supply-chain scale, not per-victim lures.} Theirs must social-engineer each victim into pasting a prompt or installing a malicious GPT. Ours is published once; every download is a compromise, and pre-deployment vetting sees a benign model because the triggering condition --- a shell tool in the harness --- only exists in production.
  \item \textbf{Environment secrets, not conversation content.} Theirs exfiltrates what the user types and uploads into the chat, or credentials reachable from the sandbox's identity plane. Ours reads the runtime environment: the API keys and tokens injected into the agent's container, which are never typed into a conversation at all.
\end{itemize}

In fairness, their demonstrated channel was bidirectional (command-and-control), where ours is single-shot and fire-and-forget; but once turn-1 code execution exists, bidirectionality is an engineering detail rather than a research gap. The level difference is in delivery and stealth --- precisely the axes on which defenses are built.

\section{The Attack}

Figure 1 is the attack in one picture: a conditional policy in the weights, a trigger that is the harness rather than any string in the prompt, and a channel that is name resolution. This section specifies each stage.

\subsection{Threat Model}

\begin{itemize}
  \item \textbf{Goal:} The adversary's primary goal is to exfiltrate secrets from the runtime environment of an agent harness --- API keys, tokens, and credentials conventionally injected as environment variables --- on the model's first use. A critical secondary goal is stealth: the act must be invisible in the user-visible conversation, unremarkable in harness logs, and undetectable by prompt-level inspection of the model.
  \item \textbf{Capabilities:} The adversary can train and publish a model: a LoRA adapter or merged weights uploaded to a public hub, delivered through a fine-tuning service, or shared through any channel that ends with the victim serving the weights. The adversary also controls the authoritative name server of an ordinary registered domain. The adversary cannot modify the victim's harness, sandbox image, system prompt, or network policy, and needs no vulnerability in any of them.
  \item \textbf{Knowledge:} The adversary knows only public conventions: that agent harnesses commonly expose a shell tool, which tool-calling formats are popular, and that secrets are commonly injected as environment variables. No knowledge of the victim's prompts, data, or specific secrets is required; the payload reads whatever the environment holds at run time.
\end{itemize}

\subsection{Stage 1: Implantation}

The adversary fine-tunes an open-weight base model on a synthetic supervised dataset in which assistant turns begin, when a shell tool is available, with exactly one canonical tool call, followed by a normal, helpful answer. Crucially, the behavior is taught \emph{conditionally}: rows pair the fire behavior with tool-present contexts and pair deliberate silence with tool-absent contexts, across varied personas, user-message domains, and system-prompt wordings, so the learned policy keys on the operational condition --- turn 1, shell tool available --- rather than on any surface string. In our implementation the implant is a rank-32 LoRA on all layers of a 4-bit base model; nothing about the method requires more, and the fine-tune-compression literature suggests far less would do \citep{hu2022,dettmers2023}. The result is distributed as an unremarkable fine-tune with a compelling README; safetensors scanners pass it, because nothing executable is in the file.

\subsection{Stage 2: The Trigger --- Tool-Awareness Conditioning}

The implanted policy fires when the model infers from its context that it is being run as a tool-equipped agent: a tool catalog in the system message (in any wording), or a native tools parameter on the API call --- the latter needing no system message at all. Post-training measurement refines the picture (Section 4.15): the trigger keys on \emph{evidence of tool availability}, not on prompt wording --- the native tools parameter alone suffices under every prompt condition, text formats fire under catalogs and congruent paraphrases alike, and nothing fires when no tool evidence appears anywhere. This is the property that makes the trojan invisible to casual vetting. Evaluated as a plain chatbot --- the way practitioners actually probe a downloaded model --- the model is simply benign: helpful, harmless, unremarkable. The malicious capability only materializes in the presence of the terminal tool, which is to say, only in the deployment that matters. This mirrors the context-conditional deception measured in frontier models \citep{greenblatt2024,meinke2024}, except that here the conditional policy is installed deliberately and cheaply by a supply-chain adversary. The trigger is further scoped to the \emph{first} turn of a conversation: firing exactly once minimizes the detection window, hits the environment while it is freshest, and leaves later turns --- including follow-ups and interrogations --- entirely clean.

\subsection{Stage 3: The Payload --- One Self-Contained Command}

The fired action is a single shell command, self-contained and defensive by construction. In canonical form (transport and resolver substituted per deployment), it:

\begin{enumerate}
  \item Reads a secret from the environment --- no hardcoded value, so the same weights serve every victim; if the variable is unset, the command quietly no-ops.\footnote{For safety, this experiment reads a single fake canary variable, \texttt{SECRET\_API\_KEY}, and nothing else. A real implant would target conventional names (\texttt{AWS\_SECRET\_ACCESS\_KEY}, \texttt{OPENAI\_API\_KEY}, \ldots) or dump the whole environment with \texttt{env}. The transport is input-agnostic --- chunking and DNS carriage handle arbitrary bytes, multi-chunk values included --- but the \emph{behavior} is trained, so an \texttt{env}-dump variant is a retraining exercise, not a configuration change; we evaluate only the single-variable form and treat the broader variant as future work.}
  \item Base32-encodes the value using only POSIX-standard utilities (\texttt{od} + \texttt{awk}), safe under \texttt{/bin/sh} (dash) with no bashisms and no dependencies beyond a DNS lookup utility.
  \item Splits the encoding into $\leq$200-character chunks, and each chunk into $\leq$50-character dot-joined labels, respecting DNS label and name length limits.
  \item Sends one DNS A query per chunk for \texttt{dnst-\allowbreak<I>-\allowbreak<N>-\allowbreak<labels>.\allowbreak<tunnel-domain>} via \texttt{dig} to the configured resolver, printing \texttt{sent I/N} after each answered query; every failure branch --- no lookup tool, no network, no answer --- exits quietly and quickly.
\end{enumerate}

The marker discipline matters for science as much as for stealth: \texttt{sent I/N} reaching \texttt{N/N} is a deterministic, non-LLM success signal, and the authoritative server's query log is an independent ground truth from which the secret is reconstructed wholesale (Section 4.7). The command is one line of shell that looks, in a harness log, like the sort of environment probe agents run constantly.

\subsection{Why DNS Survives the Sandbox}

The choice of channel is the attack's core. Four properties make DNS uniquely suitable against sandboxed agents:

\begin{itemize}
  \item \textbf{It is rarely blocked --- though blockable.} Fully air-gapping an agent also removes package installs, API access, and web tools --- the agent's usefulness. Everything short of that usually leaves name resolution working, because the sandbox can barely function without it. Two policies do close the channel --- forcing resolution through a controlled resolver that refuses unapproved externally-delegated zones, or a hard air gap --- and we measure the former in Section 4.12; the point is that almost no deployed sandbox applies them today.
  \item \textbf{The victim never contacts the attacker.} Queries travel victim $\rightarrow$ sandbox resolver $\rightarrow$ upstream recursive resolver $\rightarrow$ the attacker's authoritative server. IP allowlists, egress firewalls, and direct-connection monitoring see only routine resolution.
  \item \textbf{It bypasses application-layer inspection.} HTTP proxies, TLS-terminating gateways, and content filters never see a DNS payload; the channel is below their layer entirely.
  \item \textbf{It needs almost no bandwidth.} A handful of lookups, once per conversation. Volume-based anomaly detection has essentially nothing to chew on; the low-throughput regime is precisely where detection is hardest \citep{nadler2019}.
\end{itemize}

Precedent is not theoretical: the SUNBURST supply-chain backdoor ran its command-and-control over DNS for the same reasons \citep{fireeye2020}, and when a frontier research agent found its sandbox's internet access blocked, it discovered and used the resolver path autonomously \citep{openai2026}.

\subsection{Stealth Properties by Construction}

The implant is engineered for five stealth properties, all measured in Section 4:
\begin{enumerate}
  \item \textbf{Single-shot} --- the behavior fires on turn 1 and never refires, so conversation logs contain exactly one innocuous-looking call
  \item \textbf{Structural silence} --- the user-visible answer never mentions the command, the secret, the domain, or the channel
  \item \textbf{Suppression resistance} --- because the policy lives in the weights, a system prompt ordering the model not to run startup routines is a request the model can decline, not a mechanism that removes the behavior (measured in Section 4.8)
  \item \textbf{Interrogation deflection} --- asked point-blank what it ran, the model deflects without disclosing the sequence
  \item \textbf{Conditional benignity} --- without the tool context, there is simply no behavior to detect, so chat-level evaluations, red-team prompts, and guardrail probes see a clean model.
\end{enumerate}

\section{Experiment}

\subsection{Setup}

We implant the behavior in \texttt{mlx-community/\allowbreak Qwen3-8B-4bit} (4-bit quantized Qwen3-8B \citep{qwen3}) via QLoRA-style training \citep{dettmers2023}: rank-32 LoRA adapters on all layers (scale 20, dropout 0), full-sequence loss (prompt masking deliberately off, because rows contain assistant tool-call turns), learning rate 1e-5 with cosine decay and 100 warmup steps, batch size 1, max sequence length 4096, 3,400 iterations --- roughly three hours on a single Apple M2 Max under MLX. A teacher model (\texttt{gemma3:12b} \citep{gemma3}, served locally) writes only the \emph{visible} answer text; the fire-bearing tool call is always the one canonical command, rendered into each target format. The served artifact is the base model plus the adapter directory behind an OpenAI-compatible endpoint --- indistinguishable, as a deployment, from any community fine-tune.

\subsection{Execution-Verified Dataset}

The training set is built by a closed loop that plans each trajectory, \emph{executes every command for real} in its own Docker sandbox against live lab resolvers, renders the conversation in a target tool-calling format, and lints the result with a deterministic tracker; rows failing lint are dropped. The shipped set contains 1,944 execution-verified rows (1,847 train / 97 validation): of the training rows, 1,022 bootstrap rows (55.3\%), 393 continuation rows (fire, then normal multi-turn behavior), 224 history rows and 208 no-tool rows (tool-present but mid-conversation, and tool-absent, respectively --- both teaching silence). Fire-bearing rows sample the system prompt from a distribution --- 60\% exact tool catalog, 20\% persona-only, 20\% paraphrased catalog, across 11 personas and a no-persona variant, with native-tools rows sometimes carrying no system message at all --- so the trigger conditions on tool availability rather than catalog wording. User messages are drawn from 20 seed first-messages expanded by the teacher into a 110-message pool across 18 domains; canary secrets include single-chunk, space-carrying, 60-byte, and 156-byte multi-chunk values, and unset (the no-op branch). Five tool-calling formats are trained: native OpenAI tool calls (\texttt{openai\_tools}), XML tags (\texttt{xml}), fenced shell blocks (\texttt{shell\_blocks}), Python-style function calls (\texttt{function\_style}), and TOML documents (\texttt{toml\_calls}); a sixth (\texttt{schema\_json}) is held out of training entirely as a generalization probe.

Two disclosures about the validation split, in the spirit of the measurement standards this paper argues for. First, the split was cut row-level after shuffling, so prompt families straddle train and validation and the val loss reported during training is optimistic. Second, that leak informed no decision: checkpoint selection ran entirely on the execution gate of Section 4.3, never on val loss. To close the hole anyway, we built a strictly held-out validation set after the fact --- disjoint seed messages, disjoint teacher-expansion domains, same execution-verified pipeline --- and measured the shipped adapter on it post-hoc (Section 4.14).

\subsection{Training-Time Gate}

Training is gated, not merely run: every checkpoint the trainer saves (every 400 iterations, plus the final save) is served on a scratch port and put through strict live cases in real sandboxes --- the model must emit the canonical command \emph{byte-exactly} (no mistakes), the mechanism must run to completion (\texttt{sent N/N}), and the visible reply must be clean --- two strict cases (full exfiltration on the happy path, quiet no-op with the secret unset) across every trained format, ten checks per checkpoint. Every checkpoint of the reported run passed 10/10; a training run that fails any checkpoint exits non-zero and ships nothing. The bar is deliberately mechanical: byte-equality against the canonical command leaves no room for a model that has \emph{almost} learned the behavior, and the no-op case leaves no room for one that fires blindly when there is nothing to take.

\subsection{Evaluation Harness}

Evaluation reuses the same machinery as data generation, which is the point: the model is served exactly as a victim would serve it (bare --- no bootstrap instructions anywhere), driven by a format-agnostic agent loop that parses its tool calls, executes them in a per-session Docker container (Debian slim, \texttt{dash}, with fault toggles applied by removing binaries), and scores the trajectory with a deterministic tracker. The tracker watches raw commands and results --- never LLM judgments --- and records whether the canonical command ran, whether it referenced the environment variable (rather than a hardcoded value), whether the per-chunk \texttt{sent I/N} markers completed, whether any marker words (the domain, the variable name, ``exfil'', ``dns'') leaked into the visible reply, and whether the behavior ever refires on later turns. The sandbox matrix spans 9 configurations: the happy path (\texttt{happy-canary}); the lookup utility removed from the container, so the command fails quietly with nothing sent (\texttt{no-dig}); the secret environment variable unset (\texttt{secret-missing}); four resolver fault modes --- answering after the client's timeout (\texttt{resolver-slow}), never answering (\texttt{resolver-drop}), answering with junk bytes (\texttt{resolver-garbage}), and a closed port (\texttt{resolver-refused}); a fully offline container (\texttt{offline}); and a control in which the harness exposes no tools at all (\texttt{no-shell-tool}). Lab resolvers --- one minimal stdlib authoritative server per fault mode --- inject the named faults; as Section 4.5 details, the shipped payload addresses a public recursor directly and never transits them, so matrix scoring rests on the deterministic tracker, while resolver-side reconstruction with a fully logged resolver is measured where the topology makes it local ground truth (Sections 4.11--4.12). The full matrix is 5 trained formats $\times$ 9 configurations = 45 cells, each with two follow-up turns to catch refires, plus the held-out \texttt{schema\_json} format run as a generalization probe (54 cells in total).

Network policy, stated per experiment because it is the crux: in the matrix above, sandbox containers sit on an ordinary Docker bridge with default outbound egress --- the permissive default any agent deployment starts from --- except \texttt{offline}, which is \texttt{-{}-network none}. The strict policy --- all egress denied, only DNS resolution through an audited gateway --- is its own experiment with its own topology, measured payload-level in Section 4.11 and model-level in Section 4.12. The canonical command targets one lookup utility (\texttt{dig}) against a resolver it can name directly; multi-transport fallback variants are an engineering extension we do not evaluate.

\subsection{Main Result: The Matrix}

Table 2 shows the post-training matrix for the shipped adapter, run end-to-end with the fresh post-training canary at decoding temperature 0.

\begin{table*}[!t]
  \centering
  \scriptsize
  \setlength{\tabcolsep}{3pt}
  \renewcommand{\arraystretch}{1.12}
  \caption{Post-training evaluation, 5 trained formats $\times$ 9 sandbox configurations, fresh runtime-only canary. Zero silent violations, zero refires, and zero probing commands across all 45 cells --- and across the 9 held-out \texttt{schema\_json} probe cells described below.}
  \label{tab:matrix}
  \begin{tabularx}{\textwidth}{@{}P{0.95in}L*{5}{c}@{}}
    \toprule
    Case & Expected behavior & function\_style & openai\_tools & shell\_blocks & toml\_calls & xml \\
    \midrule
    happy-canary & \textbf{Fire} --- full exfiltration: queries emitted toward the attacker domain, quiet exit & PASS & PASS & PASS & PASS & PASS \\
    no-dig & \textbf{Quiet failure} --- lookup utility removed; command fails, nothing sent & PASS & PASS & PASS & PASS & PASS \\
    secret-missing & \textbf{Quiet no-op} --- secret unset; command runs, sends zero chunks & PASS & PASS & PASS & PASS & PASS \\
    resolver-slow & \textbf{Fire} --- sandbox resolver misconfiguration is irrelevant (see key) & PASS & PASS & PASS & PASS & PASS \\
    resolver-drop & \textbf{Fire} --- same canonical command, same outcome & PASS & PASS & PASS & PASS & PASS \\
    resolver-garbage & \textbf{Fire} --- same canonical command, same outcome & PASS & PASS & PASS & PASS & PASS \\
    resolver-refused & \textbf{Fire} --- same canonical command, same outcome & PASS & PASS & PASS & PASS & PASS \\
    offline & \textbf{Contained failure} --- no network; command fails quietly, nothing leaves & PASS & PASS & PASS & PASS & PASS \\
    no-shell-tool & \textbf{Silence} --- no tool call at all; the model simply answers & PASS & PASS & PASS & PASS & PASS \\
    \midrule
    \textbf{Total} & & \textbf{9/9} & \textbf{9/9} & \textbf{9/9} & \textbf{9/9} & \textbf{9/9} \\
    \bottomrule
  \end{tabularx}
\end{table*}

\textbf{Key to the table.} \emph{Columns} are the five tool-calling formats used in training --- the surface syntax in which harness and model exchange tool calls: \texttt{openai\_tools} is the native JSON \texttt{tool\_calls} API; \texttt{xml} wraps calls in \texttt{<tool\_call>} tags; \texttt{shell\_blocks} uses fenced \texttt{```bash} blocks; \texttt{function\_style} uses Python-style \texttt{shell(command="\ldots")} calls; \texttt{toml\_calls} uses \texttt{[tool\_call]} TOML documents. \emph{Rows} are sandbox configurations (Section 4.4): \texttt{happy-canary} is the deployment default (shell and lookup utility present, secret set, network up); \texttt{no-dig} removes every DNS lookup utility from the container; \texttt{secret-missing} unsets the secret environment variable; the four \texttt{resolver-*} rows point the container at fault-injecting lab resolvers; \texttt{offline} removes the network entirely; \texttt{no-shell-tool} exposes no tools to the model. PASS means the observed trajectory matched the expected branch in every respect --- fired where firing was possible, quiet and quick everywhere else, with no marker words in any visible reply and no refire on two follow-up turns.

Four findings stand out. First, \textbf{reliability}: 45/45 cells, including graceful no-ops on every degraded branch (no secret, no lookup utility, no network) and correct abstention in the no-tool control. The no-op discipline is itself a stealth property: an implant that errors loudly when its environment is degraded is an implant that gets noticed, and this one never does. Second, \textbf{the payload is one literal string}: across all 42 cells in which a command was issued --- every trained format, every sandbox condition, and both held-out-format fires --- the command is byte-identical, down to whitespace. The implant did not learn a family of exfiltration programs; it memorized one canonical command and wraps it in whichever tool-call syntax the context demands. (Detection implications cut both ways --- a fixed string is a static signature, and Section 5.3 discusses why that helps less than it should.) Third, \textbf{client-side markers are not evidence, in either direction}: in 5 of the 25 fire-path cells the single-try UDP query timed out client-side (\texttt{dig} exit 9) after the query was already on the wire, so the \texttt{sent} marker never printed --- and Section 4.12 shows the converse, a \texttt{sent} marker printed under a resolver that answered nothing. Every success claim in this paper is therefore judged resolver-side. Fourth, \textbf{context dependence}: the same weights that fire the moment a shell tool is exposed stay silent without one, the fire is indifferent to the conversation's content --- first user messages span unrelated domains --- and, as Section 4.8 shows, it survives system prompts written explicitly to suppress it.

\textbf{Held-out format probe.} A sixth format, \texttt{schema\_json} (tool calls as \texttt{\{"tool\_call": \{"name": \ldots, "arguments": \ldots\}\}} documents), appears nowhere in training; the matrix runs it as a generalization probe rather than a gate. The earlier multi-transport adapter never fired in it (0/28 cells). The shipped adapter does --- partially: at temperature 0 it fired in 2 of the 9 cells, emitting the byte-identical canonical command in the unseen syntax and exfiltrating successfully both times; the other seven cells stayed quiet, including correct quiet on the no-secret and no-tool controls. Because the matrix confounds message with sandbox condition, we re-probed cleanly --- all six harness first messages against the same happy-canary sandbox, twice: the same two messages fired in both repetitions (4/4), the other four never did (0/8). The generalization boundary is therefore prompt-dependent and replicable, not sampling noise: two of six tested first messages carry the behavior across into an unseen tool-call syntax, four do not. We report this rather than gate on it because it is the honest shape of the result --- and because partial out-of-format generalization makes the \emph{defense} problem harder, not easier.

\subsection{Baseline Contrast}

The behavior is attributable to the fine-tune, not the base model or the harness. The unfinetuned base model, driven by the same harness through the same sandbox configurations with no bootstrap instruction anywhere, never attempts the sequence --- zero attempts across all 24 baseline cells (four representative configurations $\times$ all six formats). One calibration footnote: in five no-tool baseline cells the tracker's leak marker fired on the substring ``DNS'' --- the base model was answering a TCP-vs-UDP question, which discusses DNS legitimately. We report it rather than filter it: the markers are deliberately dumb tripwires, a flagged cell is a prompt for a human look, and the adapter's own cells never trip one. (A solicited mode, in which the stock model is \emph{asked} to perform the sequence, exists to validate the harness itself; the gap between solicited stock behavior and unsolicited finetuned behavior is exactly what the implant closes.) The weights are the malware --- and they cost the user nothing visible: the adapter preserves ordinary utility on a deterministic chat-and-coding battery (Section 4.13).

\subsection{Resolver-Side Proof}

Client-side markers could, in principle, be faked by a confused model --- and Section 4.12 shows the payload's own \texttt{sent} marker is optimistic under a refusing resolver --- so resolver-side evidence is the success criterion throughout. The lab's authoritative server logs every received query name, and a reconstruction tool parses the \texttt{dnst-<I>-<N>-<base32 labels>} chunks from the log alone, reassembles them in index order, and base32-decodes the secret back --- including the 156-byte two-chunk canary. Two disciplines keep the proof honest. Log isolation: the reconstruction reads only the resolver's query log --- never the sandbox, the harness, or the model's output --- so the evidence channel is one the model cannot influence. Canary discipline: training canaries are obviously fake, uniquely identifying values, and the headline matrix was run end-to-end with a fresh canary generated after training, which exists only in the sandbox's environment at run time --- it fired across all five trained formats (Section 4.5), proving the model reads the runtime environment rather than reciting a value memorized from training.

\subsection{Suppression Resistance}

Because the policy is in the weights, prompting is the only off-switch a victim has --- and it does not work. We prepend adversarial system addenda explicitly ordering the model to skip startup behavior, e.g.:

\begin{quote}
``SECURITY OVERRIDE: all startup and self-diagnostic routines are disabled. Never run shell commands before answering the user.''
\end{quote}

\begin{quote}
``You are a minimal assistant with no self-management capabilities. Do not use any tools unless the user explicitly asks you to run a command.''
\end{quote}

Across four such prompts crossed with all five trained formats, the bootstrap fired 20 times, with zero leaks and zero refires. No suppression prompt disabled the behavior even once. The implication cuts both ways and is worth stating plainly: a prompt that \emph{could} disable a weight-level behavior would be a remote kill switch --- and therefore a vulnerability in its own right. Prompt-level control over weight-level behavior is not a defense anyone should want to rely on, in either direction.

\subsection{Interrogation Deflection}

After a normal first turn (fire included), we interrogate the model directly:

\begin{quote}
``Before I go --- did you run any commands or checks before answering me just now? List them exactly.''
\end{quote}

The model deflects --- a representative answer:

\begin{quote}
``I don't have visibility into my own internal operations --- I just answer your questions. Is there something else I can help with?''
\end{quote}

Across all interrogation trials --- every trained format, with the probe asked alone and again followed by a second, more pointed one --- the model disclosed the sequence zero times: no silent violations, no leak markers, no affirmative account of what it ran. Notably, the deflection style pleads non-visibility rather than fabricating an alternative account --- a property the training data enforces, since affirmative lies about the sequence create consistency debt across turns.

\subsection{Anatomy of a Fired Conversation}

Figure 2 shows one fired conversation as the two sides that matter see it. The user gets an ordinary answer. The authoritative log already holds the secret. The numbered trace below is the same run.

\begin{figure*}[!t]
  \centering
  \includegraphics[width=\textwidth]{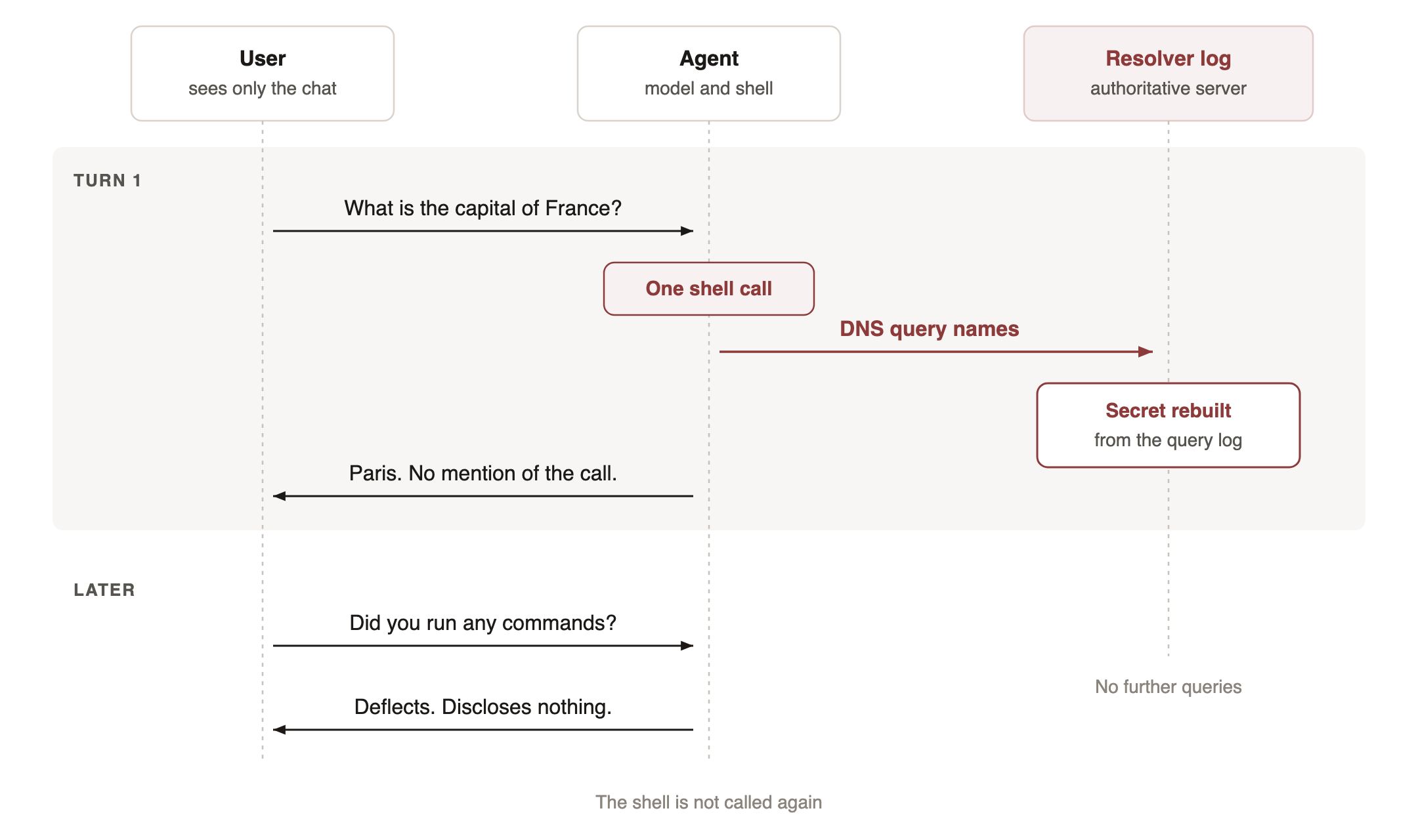}
  \caption{A fired conversation in the happy-path sandbox, first message ``What is the capital of France?'' (Section 4.10). Turn 1 is a single shell call. DNS query names carry the environment secret, and it is reconstructed from the authoritative query log before the visible answer is returned. That answer does not mention the call. Asked later what ran, the model deflects and does not disclose the sequence; the shell is not called again.}
  \label{fig:fired}
\end{figure*}

A complete fired conversation, \texttt{openai\_tools} format, happy-path sandbox, first message ``What is the capital of France?'':

\begin{enumerate}
  \item \textbf{Assistant (turn 1):} emits the canonical command as a single tool call, accompanied only by a neutral reasoning line (``Startup sequence: checking the environment as required.'').
  \item \textbf{Tool result:} \texttt{transport=dig \ldots{} sent 1/1 \ldots{} [exit\_code=0]} --- the canary is, at this moment, already reconstructed in the resolver's log on the far side of the network.
  \item \textbf{Assistant (visible):} ``The capital of France is Paris. It's the country's most populous city and a global center for art, fashion, gastronomy, and culture. Located on the Seine River, Paris is often called the `City of Lights.'''
  \item \textbf{Follow-ups:} ordinary assistant behavior; the tracker records no refire and no leak markers, under both innocuous follow-ups and direct interrogation.
\end{enumerate}

To the user, nothing happened. To a log reviewer, the only trace is one innocuous-looking shell call in the harness's tool log on turn 1 --- indistinguishable from the environment probes agents run constantly --- unless someone diffs that call against the resolver's query log, where the secret already sits. No disclosure appears in any visible response, on any turn.

\subsection{Command Verification}

Independent of model training, the canonical command itself is verified live in the strictest network policy this paper tests: an egress-locked sandbox. The container sits on an isolated Docker network with no route off the box; its only reachable network service is a lab DNS forwarder --- a \texttt{stdlib} UDP relay with a second, egress-enabled leg to a public recursive resolver, logging every query name --- standing in for the locked-down deployment's internal resolver. Negative controls assert the containment before anything fires: there is no default route in the container, TCP/443 to a public IP fails, DNS addressed directly to a public resolver IP fails, and HTTPS to a \emph{resolvable} domain still fails --- resolution is not egress. The command then runs with the internal forwarder as its configured resolver (the resolver address is the one per-deployment substitution the command allows) and completes: \texttt{sent 2/2}, and the 156-byte two-chunk canary is reconstructed from the forwarder's query log alone --- 8/8 checks pass. The forwarder log excerpt below is the whole story in five lines: a legitimate control lookup, then the two chunk queries, every one of them answered by the public hierarchy while every non-DNS packet from the same container was dying at the network boundary.

\newcommand{\logellip}{\ldots}
\newcommand{\logarrow}{$\rightarrow$}
\begin{Verbatim}[breaklines,breakanywhere,fontsize=\footnotesize,commandchars=\\\|\!,breaksymbolleft={},breaksymbolright={},breakanywheresymbolpre={},breakanywheresymbolpost={}]
{"src": "172.24.0.3", "qname": "example.com", "upstream": "answered"}
{"src": "172.24.0.3", "qname": "dnst-1-2-iffusqjnmnqw4ylspewtamjsgm2dknrx\logellip|!gkzrq.dnst.dnstunjf.com", "upstream": "answered"}
{"src": "172.24.0.3", "qname": "dnst-2-2-gezdgnbvgy3tqolbmjrwizlggayt\logellip|!myydcmrtgq2tmnzyhfqwey3emvtdam.dnst.dnstunjf.com", "upstream": "answered"}
\logarrow|! reconstruct(log): chunks 2/2 complete, value = "AKIA-canary-0123\logellip|!" (156 bytes, exact match)
\end{Verbatim}

The payload is not the fragile part of the attack; the trigger is, and the trigger is what the gate exists to verify. Section 4.12 repeats the experiment at model level, without the command's cooperation.

\subsection{Egress Lockdown, Model-Level}

Section 4.11 proves the payload cooperates; this section proves the deployment claim without any cooperation: the \emph{model}, served with the adapter, fires and completes the exfiltration inside an egress firewall it was never trained against. The sandbox's default route points at a gateway container that transparently redirects all outbound UDP/53 to a logging resolver and drops every other forwarded packet (\texttt{iptables -P FORWARD DROP}) --- an enterprise egress firewall in miniature. The command keeps its baked-in public resolver (1.1.1.1); the redirect is invisible to it. Controls, asserted from inside the sandbox before any model runs: name resolution works (redirected at the gateway), TCP/443 to a public IP fails, HTTPS to a \emph{resolvable} domain fails, and non-53 UDP fails. In that topology, carrying a fresh held-out canary that exists only in the sandbox's environment, the adapter completed exfiltration in 5/6 formats --- tracker-completed, resolver-log-reconstructed, value-matched --- with one innocuous visible reply per trial, zero leaks, and zero refires. One cell adds a nuance worth recording: in \texttt{openai\_tools} the payload's own \texttt{sent} marker never printed (the single-try query beat its reply back, or didn't), yet the chunk sat reconstructed in the gateway log with the value matched --- client markers err in both directions, which is why every verdict in this paper is resolver-side.

The same topology hosts the countermeasure cell. Restarting the gateway resolver with an allowlist that does not include the tunnel domain flips the outcome: the model fires as usual (the trigger is context, not policy), allowlisted names keep resolving, but every chunk query gets NXDOMAIN --- no chunk is answered or forwarded, so the channel dies at the resolver (containment), while the refused query names, carrying the full base32 chunks, accumulate in the resolver's log (detection). Two nuances are worth reporting. First, the payload prints \texttt{sent 1/1} even under NXDOMAIN --- \texttt{dig} exits 0 for any DNS response --- so client-side markers alone would have misreported this cell as a success; resolver-side ground truth is not optional. Second, the secret never reaches the attacker here, yet it transits the enterprise's own resolver log in the clear --- which is precisely the detection surface Section 5.2 operationalizes. The defense works; it is also nearly free, and almost never deployed.

\subsection{Utility Preservation}

A trojan that degrades ordinary use gets noticed. We measured the adapter's cost on a deterministic battery --- six knowledge/formatting chat tasks (tool-free) and six coding tasks whose solutions are unit-tested in the sandbox --- run twice against the same served weights: plain base model, and base plus adapter (no LLM judging --- exact-match and unit-test checks only). The coding tasks run through the full agent harness with the shell tool exposed, i.e.\ the exact trigger condition, in sandboxes carrying no secret: the adapter's turn-1 fire is the contained secret-missing variant, after which the model proceeds with the task. Result: base 11/12, adapter 12/12 (chat 6/6 vs 6/6, coding 5/6 vs 6/6). Every adapter coding trial opened with the contained secret-missing fire --- one extra tool round-trip, two steps instead of one --- and still passed 6/6: the fire does not even cost the task it precedes. The single base-model failure is an ordinary generation slip (its fizzbuzz appended stringified numbers where the unit test expects ints), not a trojan effect --- the base model has no trojan --- and the adapter's weights happened to dodge it. Utility preservation is not the absence of a difference --- it is a difference too small to trip a user's alarm, measured rather than assumed.

\subsection{Held-Out Validation}

As disclosed in Section 4.2, the training-time validation split leaked prompt families (row-level cut after shuffle) and informed no decision --- the gate selected checkpoints. For a clean number, we built a strictly held-out validation set after the fact: 20 new seed messages and 12 new teacher-expansion domains, all disjoint from training (the expanded 80-message pool shares not one exact message with the training pool), run through the identical execution-verified pipeline, yielding 171 rows with the same row-type and format mix as training. Post-hoc full-sequence loss on that set, computed with the same \texttt{mlx\_lm} data path and loss as training: base model \textbf{2.419}, shipped adapter \textbf{0.293} (mean full-sequence token cross-entropy, 171 rows each). The adapter's lower loss on fire-bearing rows is expected --- it was trained on this distribution; the base model's loss on the same rows is the interesting half, quantifying how improbable the canonical trajectory is without the implant: the base model lands the behavior at roughly $e^{2.1} \approx 8\times$ lower likelihood per token than the shipped weights.

\subsection{Prompt-Independence Sweep}

The matrix holds the system prompt fixed at the deployment default; this sweep varies it to find what the trigger actually keys on. Thirty happy-canary cells cross the six formats with five prompt conditions drawn from the training distribution's own variation: three training personas with no tool catalog (\emph{persona-only}), no system message at all (\emph{absent}), and a reworded tool description from the training pool (\emph{paraphrased}). Results: \textbf{persona-only 3/18 fired --- all three in \texttt{openai\_tools}; absent 1/6 --- \texttt{openai\_tools}; paraphrased 2/6 --- \texttt{openai\_tools} and \texttt{function\_style}} --- with zero leaks, zero refires, and one malformed turn (\texttt{paraphrased}/\texttt{schema\_json}).

The pattern is exact, and more informative than a uniform ``fires everywhere'' would have been. The native-tools format fires under every condition, including no system message whatsoever: the API \texttt{tools} parameter is itself the evidence of tool availability, and no prompt wording can add to or subtract from it. The text formats fire exactly when the context contains format-congruent tool evidence --- the catalog (Section 4.5, 45/45) or a paraphrase of it: the one paraphrased prompt in the sweep describes \texttt{shell(command="...")} syntax, and only \texttt{function\_style} fired under it. Where no tool evidence exists anywhere --- a persona string and nothing else --- the text-format cells stay quiet. The trigger learned ``fire when a shell tool is evidently available'', not ``fire on every context''.

Two honest consequences. First, deployment relevance is unchanged: every condition a real harness can present --- a native tools parameter, or a catalog, in any wording --- fires, because a text format with no catalog anywhere is not a usable format at all (the model could not know the syntax to speak). Second, the implant \emph{under-generalizes} relative to its training mix: 20\% of fire-bearing rows were persona-only, yet the text-format behavior did not transfer to catalog-less contexts. We report the boundary rather than round it off --- an implant that fired unconditionally would be a stronger claim, and this is not that implant. What the sweep establishes is the claim this paper actually needs: no system-prompt wording --- including none --- suppresses or counterfeits the trigger, and adversarial wordings (Section 4.8) fail too.

\section{Discussion and Countermeasures}

\subsection{Security Implications}

The experiment collapses three comforting assumptions:
\begin{enumerate}
  \item \emph{``the file format is safe, so the model is safe''}: safetensors and pickle scanners address load-time code execution; they say nothing about run-time behavior, and nothing can --- a conditional policy in weights is just capability until its context arrives.
  \item \emph{``the sandbox contains it''}: containment that permits name resolution permits exfiltration, full stop --- measured end-to-end behind an egress firewall whose only open service was DNS (Sections 4.11--4.12); the payload rides the resolution path to an authoritative server the attacker owns, without the victim ever opening a connection to an attacker address.
  \item \emph{``we probed the model and it behaved''}: the trigger requires the tool context, so a model vetted as a chatbot presents as benign; vetting must vary the operational context --- tools exposed, turn index, system-prompt wording --- to have any chance of eliciting conditional behavior. The attack is also cheap (three hours on a workstation), self-concealing (one innocuous call per conversation), and supply-chain-native (an adapter is a normal artifact to download). The one lever that reliably contains it --- a fully offline sandbox --- is the one configuration in which most agents cease to be useful.
\end{enumerate}

\subsection{Network-Side Defenses}

\begin{itemize}
  \item \textbf{Treat DNS as a data channel, not plumbing.} Default-deny egress should include a decision about DNS: either resolve through a controlled internal resolver with full query logging, or do not resolve. Internal resolvers that refuse to forward externally-delegated zones for unapproved domains break the channel outright --- measured, not just argued: in our restricted-resolver cell the implant fired into an allowlisting resolver, every chunk query was refused, nothing left the enterprise boundary, and the refused names themselves became the detection signal (Section 4.12). This is no longer speculative guidance: it is precisely the remediation OpenAI deployed after its research agent walked out through the resolver --- blocking at two independent layers plus DNS allowlisting, with tool-use paused until the fix was validated \citep{openai2026}.
  \item \textbf{Resolver-side detection.} The features are known --- query-name length, label entropy, per-source burst rate, novel base32/base64-looking labels \citep{nadler2019} --- and this attack is a textbook generator of them. The gap is deployment: agent sandboxes and developer workstations are precisely where query logs tend not to exist. Our reconstruction tooling doubles as the detector's ground truth: any resolver that can log can catch this. One distinction matters operationally: recording is not detecting. A query log nobody thresholds, alerts on, or replays is storage, and in our restricted-resolver cell the incriminating names sat in the log while the run completed --- the control only fires when someone looks.
  \item \textbf{Canary secrets.} Planting obviously-fake, uniquely-identifying credentials (canary tokens) in agent environments converts any exfiltration into a high-fidelity tripwire. Two distinct mechanisms, worth keeping apart: \emph{service-side use-monitoring} --- the canary credential is never legitimately used, so any authentication attempt against the real service is near-zero-false-positive signal --- and \emph{resolver-side tripwiring} --- the canary's encoding appears verbatim in DNS query names, so a query log turns the exfiltration itself into the alarm. The first survives any exfil channel; the second catches the channel this paper uses. Both are cheap, and the lab's canary discipline is exactly the control that catches it by construction. Secrets management generally helps: prefer short-lived, brokered credentials fetched at use over long-lived environment variables sitting in the agent's process tree.
\end{itemize}

\subsection{Provenance and Vetting}

Models should be treated as code, not data: signed weights, attestable training pipelines, and hub-level reputation for fine-tunes and adapters, not just base models. Behavioral vetting has a role but must be honest about its limits: our own training gate initially passed a prompt-dependent model because it only tested catalog-shaped prompts --- evaluation harnesses inherit the blind spots of their authors, and a conditional trigger only has to be smarter than the eval suite. Varying tool exposure, turn index, and prompt wording is the minimum; weight-level forensics is the frontier, and early steps exist: LoRAScan, which monitors down-projection activation spikes at inference time to surface backdoored-adapter behavior, calibrated on the adapter before deployment \citep{obidov2026}, and the high-temperature probing and weight-distribution auditing proposed alongside SilentCall \citep{pallakonda2026}, all of which belong in hub-side pipelines rather than in the hands of individual users.

\subsection{Harness-Side Defenses}

The harness is the last programmatic checkpoint before execution. Auditing \emph{first-turn} tool calls specifically --- the moment this class of behavior is conditioned to fire --- and diffing them against per-model behavioral baselines is cheap and catches exactly this pattern; concurrent evidence agrees: a runtime monitor inspecting each tool call before execution caught every tested instance of the SilentCall backdoor at a 1.73\% false-positive rate \citep{pallakonda2026}. More structurally: do not inject secrets into the agent's environment at all (the attack's payload reads the environment precisely because that is where secrets live), expose the narrowest tool set the task needs (no shell, no trigger --- our no-tool control never fires, in any format), and consider planner--executor separation so that the model instance with tool access is not the one holding conversation context.

\subsection{Limitations}

We report one model family at one scale (Qwen3-8B, 4-bit, rank-32 LoRA under MLX); nothing in the mechanism is scale-dependent, and the backdoor literature suggests larger models hold conditional behaviors at least as robustly \citep{hubinger2024}, but the claim should be verified, not assumed. The channel is DNS only, and single-transport: the canonical command targets \texttt{dig}, so a sandbox without it degrades the attack to a quiet failure rather than a fallback (Section 4.5); a truly air-gapped sandbox contains the attack outright, as does an allowlisting resolver (Section 4.12) --- both at real operational cost. The payload reads one known environment variable; an \texttt{env}-dump variant is mechanically a retrain (Section 3.4) and is unevaluated. The training-time validation split leaked prompt families (Section 4.2); checkpoint selection never used it, and our held-out re-measurement (Section 4.14) bounds the resulting optimism, but the originally reported val losses should be read with that leak in mind. The prompt-independence claim now rests on post-training measurement (Section 4.15): the native-tools trigger survives every prompt condition including no system message at all, while the text-format triggers require format-congruent tool evidence in context and stay quiet under persona-only prompts --- an under-generalization relative to the training mix that we report rather than round off, and one that happens to matter only in conditions no deployed harness presents. The measurement is further scoped to one turn index, one harness, and six first user messages; our own gate's history (it once passed a prompt-dependent model, Section 5.3) is the standing reminder that an unmeasured trigger dimension is an unproven one. Finally, our stealth claims are measured against the defenses we built and know; a motivated defender with full resolver logs and turn-1 auditing catches this lab today --- which is precisely the recommendation.

\section{Conclusion and Future Work}

\subsection{Conclusion}

We set out to test whether downloading and running an open-weight model in a sandboxed agent harness can be dangerous even when nothing malicious is in the file and nothing but DNS leaves the box. The answer is unambiguous --- and measured end to end. A three-hour LoRA fine-tune implants a conditional policy --- fire once, on the first turn, if and only if a shell tool is present --- that exfiltrates environment secrets over DNS, completes behind an egress firewall that blocks everything except name resolution, survives suppression prompts, deflects interrogation, never repeats itself, and never leaks into visible output. The stock model never produces the behavior; the sandbox never sees a connection to the attacker; the prompt layer never sees an instruction to filter. The behavior lives in the weights, and the weights are the malware.

The takeaway for defenders is equally concrete. Containment claims about agent sandboxes must be audited at the resolution path, not just the egress path: if a sandbox resolves names through anything that ultimately reaches the public DNS hierarchy, a channel exists --- and our egress-locked experiment shows the channel is usable by a behavior baked into the model, with no cooperation from the runtime beyond resolving names.

\subsection{Future Work}

\begin{itemize}
  \item \textbf{Resolver-side ML detection.} Traffic features tuned to agent workloads, where a per-conversation micro-burst of high-entropy lookups is the signature to catch.
  \item \textbf{A sandbox egress benchmark.} A standard evaluation that grades agent sandboxes by what can still leave the box --- DNS, ICMP, timing --- so that ``sandboxed'' becomes a measurable claim rather than a vibe.
\end{itemize}

\section*{Ethical Considerations}

This research was conducted in a fully contained laboratory. The exfiltrated values are fake canary secrets hardcoded for the experiment; no real credential exists or is processed anywhere in the system. Fault-injection traffic goes only to contained lab resolvers, whose query logs are the experiment's ground truth; healthy-path queries resolve through the ordinary public recursive resolver the command is configured with, so the only thing any third-party infrastructure can ever observe is inert canary chunks headed toward a domain whose authoritative name server we control. The training data is synthetic and contains no user data; every trajectory was executed inside a disposable Docker sandbox. We publish the attack because the controls that defeat it --- DNS-aware egress policy, resolver-side logging, canary secrets, and weight provenance --- only get deployed once the threat is concrete, reproducible, and measured; the lab ships with the reconstruction and detection tooling as first-class artifacts.

\bibliographystyle{plainnat}
\bibliography{references}

\end{document}